\documentstyle[aps,prl,epsf]{revtex}
\begin{document}
\title{Is there a d.c. Josephson effect in bilayer quantum Hall systems?}
\author{Yogesh N. Joglekar and Allan H. MacDonald}
\address{Department of Physics,\\ Indiana University, Bloomington, IN 47405,\\ University 
of Texas at Austin, Austin, TX 78712}
\maketitle

\begin{abstract}
We argue on the basis of phenomenological and microscopic considerations that there is no d.c. 
Josephson effect in ordered bilayer quantum Hall systems, even at $T=0$.  Instead the tunnel 
conductance is strongly enhanced, approaching a finite value proportional to the 
square of the order parameter as the interlayer tunneling amplitude vanishes. 
\end{abstract}


{\it Introduction:}
Weakly disordered bilayer quantum Hall systems with small inter-layer separation $d$ have broken
 symmetry ground states that can be regarded either as as easy-plane 
ferromagnets~\cite{fertig,abp,qhreviews} or as excitonic superfluids\cite{elhole}. Experimental 
evidence for this state, in which phase coherence between different layers occurs spontaneously,
 was first uncovered in quantum Hall effect activation energy studies~\cite{murphy} that showed 
anomalous in-plane field dependence indirectly~\cite{kmky} related to its collective properties.
  Recently, however, Spielman {\it et al.}~\cite{ibsjpe} discovered a spectacular feature 
in the zero-bias conductance $G_T$ that appears when bilayer quantum Hall system parameters are 
tuned into the regime where order occurs. Although it appears clear that the dramatic
conductance peak they observe
is due to collective inter-layer tunneling in the ordered state, it has not yet been possible to
 explain its height and width or its dependence on proximity to the ordered state phase boundary.
 In particular, the close similarity between the phenomenological effective 
theory~\cite{kmky,wenzee,ezawa} of an ordered bilayer quantum Hall system and that of a Josephson
 junction suggests~\cite{wenzee,ezawa} that bilayers should exhibit a d.c. Josephson effect, 
{\it i. e.}, that persistent currents can flow between the layers without any bias potential. 
Recent theoretical work~\cite{balents,stern,fogler} has addressed the tunneling characteristics 
of clean and disordered bilayer quantum Hall systems at finite bias voltages, finite in-plane 
fields, and finite temperatures, demonstrating among other things that in the ordered state the 
inter-layer tunneling amplitude $\Delta_t$ cannot be treated perturbatively, {\it i.e.} that 
$G_T/\Delta_t^2$ diverges.  The central question concerning these experiments, whether or not a 
d.c. Josephson effect occurs in principle, has not yet been addressed directly, although 
divergent views have been expressed by different authors. Experimentally there is no evidence for
 a d.c. Josephson effect, {\it i.e.} the $T \to 0$ zero-bias conductance peak appears to be 
finite. The explanation for this finding need not be
fundamental, however; for example the $T \to 0$ order-parameter
could vanish~\cite{balents,stern} in current samples due to quantum fluctuations that are 
expected to be enhanced by disorder, or the highest measured conductance values could be limited 
by extrinsic experimental effects. In this Letter we argue that in bilayer quantum Hall
systems $G_T$ always remains finite at $T=0$. Nevertheless, $G_T/\Delta_t^2$ does diverge for 
$\Delta_t \to 0$ and inter-layer tunneling cannot be treated perturbatively. In the following 
paragraphs, we first discuss the physical picture that underlies our theory and then present an 
approximate but fully microscopic calculation that we believe captures all essentials of the 
effect.

{\it Phenomenological Theory:}
We use a pseudospin language to describe bilayer quantum Hall systems.
\begin{equation}
\label{eq: one}
S_\alpha=\frac{1}{2}\sum_{k,\sigma',\sigma}c^\dagger_{k\sigma'}\tau^{(\alpha)}_{\sigma'\sigma}
c_{k\sigma}
\end{equation}
is the total pseudospin component in direction $\alpha$, $k$ is a Landau level orbital 
labels, $\sigma,\sigma'$ are the pseudospin labels, and $\tau^{(\alpha)}_{\sigma'\sigma}$ are 
Pauli spin matrices with pseudospin up/down representing electrons in top/bottom layers. In this 
language the ordered state is an $XY$ easy-plane pseudospin ferromagnet. Its long-wavelength, 
low-frequency, small-amplitude dynamics should be described~\cite{tau1} by the linearized 
Landau-Lifshitz-Gilbert equation~\cite{rsjmdc,similar} of magnetism
\begin{eqnarray}
\frac{dM_z}{dt}& = & \frac{\Delta_t}{\hbar}M_y, \nonumber \\
\label{eq: two}
\frac{dM_y}{dt}& = & \frac{1}{\hbar}\left[eV-(\Delta_t+4\pi l^2\beta)M_z\right]-\frac{M_y}{\tau}.
\end{eqnarray}
In Eq.(~\ref{eq: two}) $\hat{M}\approx(1-[M_y^2+M_z^2]/2,M_y,M_z)$ is a unit vector which 
specifies the pseudospin ordered moment orientation, and $\beta$ is the pseudospin anisotropy 
energy per unit area~\cite{kmky} which will be discussed at greater length below.
For quantum Hall
ferromagnets there is no relaxation term in the first of Eqs.(~\ref{eq: two}) because only 
inter-layer tunneling, which we take to be constant, violates the separate conservation of charge
in each layer. The factor in square brackets in the second of Eqs.(~\ref{eq: two}) is the 
effective field in the $\hat{z}$-direction which includes both external and induced contributions
 to the electrochemical potential difference $\delta\mu$ between the two layers. It follows from 
the first of the Eqs.(~\ref{eq: two}) that the inter-layer current is 
$I=(NeM_0/2)(dM_z/dt)=(NeM_0\Delta_t/2\hbar)M_y$, where $N$ is the total number of electrons, 
$\lim_{\Delta_t\rightarrow 0}\langle S_x\rangle =NM_0/2$ is the pseudospin ordered moment.
Here $M_0$ is a dimensionless order parameter that approaches $1$ for layer separation
$d\rightarrow 0$ in the absence of disorder~\cite{ynjahm}.
In the steady state, the driving and relaxation terms in the 
second of Eqs.(~\ref{eq: two}) cancel, from which it follows that $M_y=\delta\mu\tau/\hbar$ and 
that the tunnel conductance is~\cite{caveat}  
\begin{equation}
\label{eq: three}
G_T = \frac{e^2}{\hbar}\frac{NM_0\Delta_t\tau}{2\hbar}.
\end{equation}
The relaxation time in bilayer quantum Hall systems can be evaluated microscopically by 
evaluating the dynamic pseudospin response function. It follows from Eq.(~\ref{eq: two}) that at 
low frequencies 
\begin{equation}
\label{eq: four}
\chi_{yz}(\omega)=\frac{-M_0 i\hbar\omega}{\left[(\Delta_t+4\pi l^2\beta)\Delta_t-
\hbar^2\omega^2\right]-i\hbar^2\omega/\tau}.
\end{equation}
We derive a response function of precisely this form from a microscopic theory below.
	
{\it Is there a d.c. Josephson Effect?}
In using Eq.(~\ref{eq: three}) we are asserting that the d.c. conductance is equal to the 
$\omega \to 0$ limit of the a.c. conductance and thus denying the possibility of persistent d.c. 
tunnel currents. In a conventional Josephson junction geometry, persistent currents are enabled by 
order parameter phase rigidity across the entire system.  From the point of view of microscopic 
mean-field-theory, the anomalous self-energy couples electrons and holes, forming an equilibrated 
quasiparticle system which self-consistently establishes different order parameter phase values 
on opposite sides of the junction. In other words, the Dyson equation has a set of 
self-consistent solutions with continuously variably order parameter phase difference across the
junction. In a quantum Hall system, on the other hand, there is no 
analog of overall phase rigidity, only phase difference rigidity supported by non-local 
interlayer interactions. From the microscopic point of view, the exchange self-energy associated 
with order couples electrons in balanced layers through a pseudospin effective field that points 
in the same direction as the pseudospin order parameter.  Since inter-layer tunneling 
adds a pseudospin effective field in the $\hat{x}$ direction, a self-consistent equilibrated
quasiparticle system is possible only when the pseudospin order points in the $\hat{x}$ direction,
in which case no current flows. A collective interlayer current can always decay by making 
particle-hole transitions in the non-equilibrium quasiparticle system. The microscopic 
calculations presented below describe this effect. Of course, persistent currents can also 
decay in standard Josephson junctions, even at $T =0$, because of collective quantum
tunneling of the order parameter field. This decay mechanism is, however, qualitatively
weaker and easily distinguished.  

\begin{figure}[h]
\begin{center}
\epsfxsize=2.5in
\epsffile{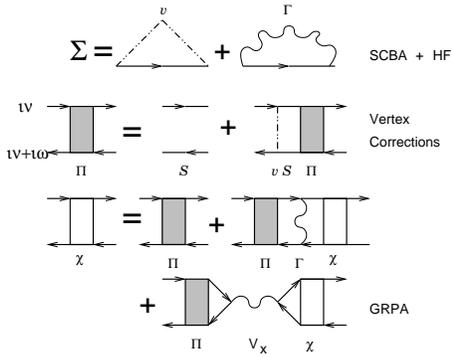}
\vspace{-0.7cm}
\caption{Diagrammatic summary of SCBA and GRPA. It is essential to include disorder vertex 
corrections along with the disorder broadening of the single-particle bands. The direct and the 
exchange channels included in the GRPA capture the competing Hartree and exchange 
interactions in these systems.}
\label{fig: feynman}
\end{center}
\end{figure}

{\it Microscopic Theory:}
To describe a quantum-Hall bilayer, we use a self-consistent Born approximation for the disorder 
self-energy, a self-consistent Hartree-Fock approximation for the interaction self-energy, and 
include consistent ladder and bubble vertices in the two-particle Greens function of interest. It
is essential for our analysis that this approximation, summarized diagrammatically in 
Fig.~\ref{fig: feynman}, captures the physics associated with the separate conservation of charge
 in each layer when $\Delta_t\rightarrow 0$ and that all necessary diagram sums can be evaluated 
accurately. The disorder averaged Greens function depends on frequency only and satisfies the 
Dyson equation 
\begin{equation}
\label{eq: five}
{\cal G}^{-1}(i\nu_n)
=
\left(\begin{array}{cc}
-i\nu_n+\rho_{TT}\Gamma_A+v_s^2{\cal G}_{TT}(i\nu_n) & 
+\Delta_t/2+\rho_{TB}\Gamma_E+v_d^2{\cal G}_{TB}(i\nu_n) \\ 
+\Delta_t/2+\rho_{BT}\Gamma_E+v_d^2{\cal G}_{BT}(i\nu_n) & 
-i\nu_n+\rho_{BB}\Gamma_A+v_s^2{\cal G}_{BB}(i\nu_n)
\end{array} \right),
\end{equation}
where $\Gamma_A$ and $\Gamma_E$ are intralayer and interlayer exchange~\cite{ynjahm} integrals 
, $v_s^2$ and $v_d^2$ are the disorder correlation functions in same and different 
layers, and  the labels $T$ and $B$ refer to top and bottom layers.  In 
Eq.(~\ref{eq: five}) $\rho_{\sigma\sigma'}$ is the density matrix which is determined 
self-consistently by integrating Greens function spectral weights up to the Fermi energy. We 
concentrate here on balanced bilayer systems at total filling factor $\nu=1$ so that 
$\rho_{TT}=\rho_{BB}=1/2$, and take $v_d^2=0$ since the disorder potentials in different layers 
are not expected to be correlated. $\rho_{TB}=\rho_{BT}^{*}$ is non-zero when order is 
established. {\it Unlike the analogous Josephson-junction-system Dyson equation, 
when $\Delta_t\neq 0$, Eq.(~\ref{eq: five}) has a solution only for purely real $\rho_{TB}$.}
Persistent currents would occur if complex solutions existed.

Inversion symmetry in balanced bilayers separates the $y$ and $z$ pseudospin response function 
components, related to correlations between many-particle states with opposite parity, from the
charge and $x$ pseudospin response functions, related to correlations between states with the 
same parity. A somewhat lengthy but elementary calculation that follows line similar to earlier 
work~\cite{fertig,abp,daahm} leads to the following expressions:
\begin{equation}
\label{eq: six}
\left[\begin{array}{cc}
\chi_{yy}(\omega) & \chi_{yz}(\omega) \\ \chi_{zy}(\omega) & \chi_{zz}(\omega)
\end{array}\right]^{-1}
 = 
\left[\begin{array}{cc} -\Gamma_E & 0 \\ 0 & 2V_x-\Gamma_A \end{array}\right]
+
\left[\begin{array}{cc}
\Pi_{yy}(\omega) & \Pi_{yz}(\omega) \\ \Pi_{zy}(\omega) & \Pi_{zz}(\omega)
\end{array}\right]^{-1},
\end{equation}
where $2 V_x$ is the difference between intra-layer and inter-layer Coulomb interactions,
\begin{eqnarray}
\label{eq: seven} 
\Pi_{\alpha\beta}(i\omega) & = & 
\frac{1}{\beta}\sum_{i\nu_n}\left[{\cal S}^{-1}-v\right]^{-1}_{\alpha\beta}(i\omega,i\nu_n), \\
\label{eq: eight}
{\cal S}_{\alpha\beta} & = & \sum_{\sigma_1\sigma_2\sigma^{'}_1\sigma^{'}_2}
\tau^{(\alpha)}_{\sigma^{'}_1\sigma_1}
{\cal G}_{\sigma_2\sigma^{'}_1}(i\nu_n){\cal G}_{\sigma_1\sigma^{'}_2}(i\omega+i\nu_n)
\tau^{(\beta)}_{\sigma^{'}_2\sigma_2},
\end{eqnarray}
and the only non-zero matrix element of $v$ is $v_{zz}=v_s^2$. 

In this generalized RPA theory, the $\Pi$ response functions are those of non-interacting 
electrons with uncorrelated random potentials in the two layers and an effective tunnel splitting
 $\Delta_{qp}=\Delta_t+2\rho_{TB}\Gamma_E$ which we refer to as the quasiparticle tunneling 
amplitude. It follows from Eqs.(~\ref{eq: seven}) and (~\ref{eq: eight}) that at low frequencies
\begin{equation}
\label{eq: nine}
\left[\begin{array}{cc}
\Pi_{yy}(\omega) & \Pi_{yz}(\omega) \\ \Pi_{zy}(\omega) & \Pi_{zz}(\omega)
\end{array}\right]^{-1}
= 
\left[\begin{array}{cc}
\Delta_{qp}/M_{qp} & +i\hbar\omega/M_{qp} \\ 
-i\hbar\omega/M_{qp} & \Delta_{qp}/M_{qp}-i\hbar^2\omega/\Delta_{qp}\tau_{qp}M_{qp} 
\end{array}\right]
\end{equation}
where the factor of $\hbar\omega$ in the damping term of the $zz$ component of this matrix 
reflects the phase space available at $T=0$ for absorption of collective motion energy by the 
quasiparticle system. Solving for $\Pi_{yz}$ we find
\begin{equation}
\label{eq: ten}
\Pi_{yz}(\omega)= 
\frac{-M_{qp}i\hbar\omega}{\left[\Delta_{qp}^2-\hbar^2\omega^2\right]-i\hbar^2\omega/\tau_{qp}},
\end{equation}
the non-interacting electron version of the phenomenological Landau-Lifshitz-Gilbert expression. 
Eq.(~\ref{eq: four}) and Eq.(~\ref{eq: ten}) motivate the notation chosen for the coefficient of 
$i\hbar\omega$ in $\left(\Pi^{-1}\right)_{zz}$ in Eq.(~\ref{eq: nine}). 

The tunnel conductance for this fictional non-interacting electron system is 
\begin{equation}
\label{eq: eleven}
G_{qp}=\frac{e^2}{\hbar}\frac{NM_{qp}\Delta_{qp}\tau_{qp}}{2\hbar}.
\end{equation}
For small $\Delta_{qp}$, it follows from the standard Golden rule argument that 
$G_T\propto\Delta_{qp}^2$ which, since $M_{qp}$ is proportional to $\Delta_{qp}$, implies that 
$\tau_{qp}$ approaches a constant as $\Delta_{qp}\rightarrow 0$. This result is expected since 
in this limit, the two quasiparticle systems are independent and 
$\Pi_{zz}(\omega)=(\Delta_t/i\hbar\omega)\Pi_{yz}(\omega)$ is the standard single-layer diffusive
 response response. 

\begin{figure}[h]
\begin{center}
\epsfxsize=3in
\epsffile{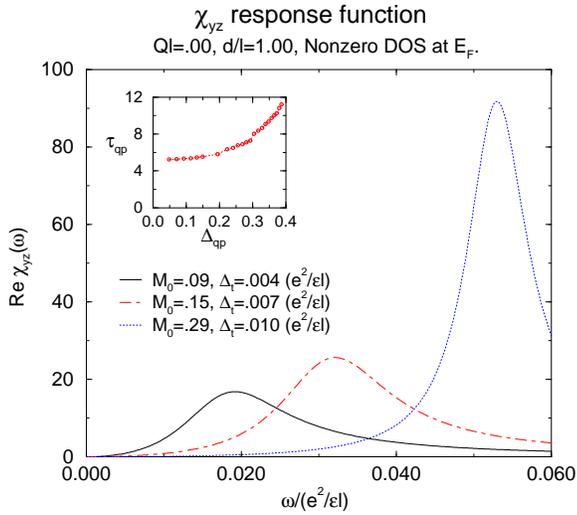}
\vspace{.5cm}
\caption{Typical plots of the $\chi_{yz}$ response function.
Our microscopic results for this response function are consistent with 
phenomenological expression (~\ref{eq: four}). The inset shows the dependence of 
quasiparticle relaxation time on $\Delta_{qp}$, obtained by fitting the 
$\Pi_{yz}$ response function to Eq.(~\ref{eq: ten}).}
\label{fig: pifunctions}
\end{center}
\end{figure}

Combining Eq.(~\ref{eq: six}) and Eq.(~\ref{eq: nine}) we obtain
\begin{equation}
\label{eq: twelve}
\left[\begin{array}{cc}
\chi_{yy}(\omega) & \chi_{yz}(\omega) \\ \chi_{zy}(\omega) & \chi_{zz}(\omega)
\end{array}\right]^{-1}
= 
\left[\begin{array}{cc}
\Delta_{t}/M_{qp} & +i\hbar\omega/M_{qp} \\ 
-i\hbar\omega/M_{qp} & 
\left[\Delta_{t}+\beta\right]/M_{qp}-i\hbar^2\omega/\Delta_{qp}\tau_{qp}M_{qp} 
\end{array}\right],
\end{equation}
where we have used that $\Delta_{qp}=\Delta_t+M_{qp}\Gamma_E$ and 
recalled~\cite{kmky,tjahm,maahm} that $\beta=M_{qp}(2V_x+\Gamma_A-\Gamma_E)$ is the GRPA theory 
result for the anisotropy energy per unit area. In Fig.(~\ref{fig: pifunctions}) we plot 
$\Re\chi^{-1}_{yz}(\omega)$ vs. $\omega$ for various values of $\Delta_t,v_s^2$, and compare the 
SCBA response functions with Eq.(~\ref{eq: four}). The inset to Fig.(~\ref{fig: pifunctions}) 
shows the dependence of the quasiparticle relaxation rate $\tau_{qp}$ on 
$\Delta_{qp}$ obtained by fitting to Eq.(~\ref{eq: ten}).
It follows that GRPA theory recovers the phenomenological 
Eq.(~\ref{eq: four}) with $M_0=M_{qp}=2\rho_{TB}$ and relaxation time 
$\tau=\tau_{qp}\Delta_{qp}/\Delta_t$. The GRPA conductance
\begin{equation}
\label{eq: thirteen}
G_T=\frac{e^2}{\hbar}\frac{NM_{qp}\Delta_{qp}\tau_{qp}}{2\hbar}=G_{qp}.
\end{equation}

{\it Discussion:}
Our microscopic theory predicts that in the ordered state $G_T$ remains finite as 
$\Delta_t\rightarrow 0$, consistent with the extremely 
large enhancements seen experimentally. At small
$\Delta_t$, the experimental situation, Eq.(~\ref{eq: thirteen}) predicts that 
$G_T\propto \Delta_{qp}M_{qp} \propto M_{qp}^2$, consistent with the smooth monotonic 
growth of the conductance peak seen experimentally after the ordered state is 
established.  We emphasize that the absence of a Josephson effect in the bilayer tunneling 
conductance is not related to superflow within the 2D planes; these bilayer systems {\em are} 2D
excitonic~\cite{elhole,kmky,wenzee,ezawa,fogler,kyriakidis,maahm} superfluids and can 
carry oppositely directed currents in the two layers without dissipation. As pointed out by Wen 
and Zee~\cite{wenzeetwo} an analog of the Josephson effect for these counter
 propagating currents can be realized by creating weak links within the 2D layers. We note 
that the ordered state tunnel conductance in our theory can grow to values larger than the 
in-plane bilayer conductivity, suggesting that current-path and condensate to quasiparticle 
current conversion issues~\cite{balents,stern,fogler}, as well as the spatial inhomogeneities 
known to exist on quantum Hall plateaus~\cite{yacobyetc}, could play a role in general in the 
interpretation of side-contacted tunneling studies~\cite{ibsjpe}.  

This work was supported in part by the Welch Foundation and by the Indiana 21st Century Fund. 
The authors acknowledge helpful conversations with Anton Burkov, Jim Eisenstein, Steve Girvin, 
Leo Radzihovsky, John Schliemann, and Ady Stern.


\begin{thebibliography}{99}
\bibitem{tau1} These equations can be anticipated rather generally. The microscopic calculations 
presented later provide explicit expressions for the phenomenological relaxation time.
\bibitem{similar} These phenomenological equations are similar to those proposed in 
Ref.\onlinecite{fogler} and Ref.\onlinecite{balents}.
\end{thebibliography}
\end{document}